# Force and energy dissipation variations in non-contact atomic force spectroscopy on composite carbon nanotube systems

A. Ilie[1], J.S. Bendall[1], O. Kubo[2], J. Sloan[3], and M.L.H. Green[3]

[1]*Nanoscience, University of Cambridge, 11 J.J. Thomson Avenue, Cambridge CB3 0FF, United Kingdom*

[2]*National Institute for Materials Science, 1-1 Namiki, Tsukuba, Ibaraki 305-0044, Japan*

[3]*Inorganic Chemistry Laboratory, University of Oxford, South Parks Road, Oxford OX1 3QR, United Kingdom*

## Abstract

UHV dynamic force and energy dissipation spectroscopy in non-contact atomic force microscopy were used to probe specific interactions with composite systems formed by encapsulating inorganic compounds inside single-walled carbon nanotubes. It is found that forces due to nano-scale van der Waals interaction can be made to decrease by combining an Ag core and a carbon nanotube shell in the Ag@SWNT system. This specific behaviour was attributed to a significantly different effective dielectric function compared to the individual constituents, evaluated using a simple core-shell optical model. Energy dissipation measurements showed that by filling dissipation increases, explained here by softening of C-C bonds resulting in a more deformable nanotube cage. Thus, filled and unfilled nanotubes can be discriminated based on force and dissipation measurements. These findings have two different implications for potential applications: tuning the effective optical properties and tuning the interaction force for molecular absorption by appropriately choosing the filling with respect to the nanotube.

## 1. Introduction

Material encapsulation [1-5] within single-walled nanotubes is one of the template methods [6] through which one can synthesize one-dimensional systems. The single-walled nanotube template constrains the topology of the encapsulated nano-crystals or nano-wires, and stabilizes them so that one can obtain atomically regulated one-dimensional structures (of just a few atomic layers in cross section) that can relate to that of the bulk material, or, on the contrary, have a completely novel structure [3]. Contrary to other template methods, such as ones using mesoporous media [7], the carbon cage (the template) does not have to be removed after synthesis to allow integration of the nano-wires in devices, while conferring protection and mechanical integrity on them, and facilitating their manipulation on surfaces. As a by-product of the encapsulation the properties of the nanotube template are also modified.

Though there has been considerable progress in diversifying the types of inorganic nano-materials that can be encapsulated in single-walled nanotubes (as, for example, semiconducting materials [8]), and towards improving their crystallinity and encapsulation yield, only very little work has been done to characterize the resulting hybrid nanotube-nanowire composites [9,10], while a few publications dealt with encapsulation in multi-walled nanotubes [11]. There are two types of information one would like to access in this regard, (i) the properties of the nano-wire itself, and (ii) the properties of the hybrid nano-composite as a whole, and from these to establish to what extent the carbon nanotube shell screens the behaviour of the nano-wire from the outside world.

The investigations presented in this paper fall into the second category. We considered the Ag@SWNT nano-composite system, and investigated how the Ag core modifies the atomic potential in the outer vicinity of the carbon cage, as probed by dynamic non-contact AFM spectroscopy in UHV. Force spectroscopy in non-contact AFM on atomically flat surfaces [12,13] demonstrated extreme resolution, being able to probe the interaction with single atomic sites [14]. However, to the best of our knowledge, this is the first time UHV force spectroscopy was used to probe specific interactions with carbon nanotube-based materials, despite quite extensive use of air AFM techniques for manipulation [15] or application of mechanical stress [16], or for electrical probing and gating [16-18]. The conditions of our measurements were such that the dominant interaction between the tip and the nano-composite was the van der Waals dispersion interaction, which is of a longer range and, thus, site independent. Based on this, the contrast observed in various types of images was interpreted as due to significant changes in the dielectric function introduced by combining materials with very different dielectric properties, such as Ag core and carbon shell. This highlights the fact that appropriately choosing the filling material would allow tuning the effective optical properties of the composite. The interaction forces with the composite can also be tuned. In this example, counter intuitively, the presence of the Ag core did not just add a component to the van der Waals interaction with the tip, by increasing the interaction force; on the contrary, this was found to decrease compared to the case of the unfilled nanotube. This conclusion could be generalized to other type of probes, such as molecules, so that filling could provide a possible way to control molecule absorption onto the composite's body. Energy dissipation [19] was recorded at the same time as the force, showing that filling increases dissipation compared to the unfilled nanotubes. This is accounted for by modifications in the elasticity of the nanotube shell, which, in this specific case, becomes more deformable (in the limit of small deformations) due to softening of C-C bonds. Thus, both force and dissipation measurements provide a means to discriminate between filled and unfilled nanotube regions. This analysis that amounts to material contrast required the *a priori* identification and separation of topographic contributions for these non-flat surfaces.

## 2. Experimental details

The Ag@SWNT nano-composites were prepared in a two-stage process. Single-walled nanotubes formed via the arc technique were first filled with high purity AgCl according to a capillary technique [20], by heating the nanotubes together with the halide. The halide filled material was then able to partially undergo light-assisted reduction to form Ag nano-wires (though AgCl remained unreduced in smaller diameter tubes). The encapsulation yield was about 60-70%, as determined by conventional high-resolution transmission electron microscopy techniques. The nano-wires were typically continuous over at least 100 nm and predominantly crystalline. The nano-composite material was then dispersed in chloroform and transferred onto a graphite substrate. Subsequent thermal annealing in UHV was used to remove residual chloroform from the substrate and the nano-composites.

A Si cantilever, of stiffness $k_c = 30$ N/m, and oscillated at resonance ($f_0 = 239$ kHz) was used for the non-contact AFM spectroscopy (in a Omicron Nanotech. dual AFM/STM microscope). The only treatment performed on the cantilever was annealing in UHV for two hours at 150 °C to remove water and other contaminants. One can thus assume that, at least initially, the end of the tip was covered with native silicon dioxide. The experimental set-up used a self-driven oscillator, whose oscillation amplitude was stabilized at about $A = 16$ nm [21]. In order to avoid artefacts, we only analysed measurements and images taken with a symmetric tip, which produced symmetric and featureless images when probing a simple cylindrical object, such as an individual single-walled nanotube. The distance range for the spectroscopic measurements was limited in order to avoid reaching the repulsive contact regime.

Because the filling yield is less than 100%, by performing single point spectroscopic measurements randomly on nanotubes there is no guarantee that we reach a part of the nanotube that has been filled. Therefore maps of simultaneous frequency-distance and dissipation-distance measurements were acquired over relatively large portions of nanotubes. For this, the tip was moved between points on a grid with the feed-back enabled, producing a certain topographic profile, and at each grid point spectroscopic curves were taken after disabling the feed-back. The resolution of the resulting maps is less than that of a normal image, for example $1 \times 1 \, nm^2$. The acquisition time (ranging between 2 to 5 s) and distance range of a spectroscopic curve were somewhat limited, in order to maintain a reasonable overall measurement time. During this procedure care was taken to minimize the drift associated with the piezo scanner. However, due to the relatively long acquisition time the measurements cannot be considered as site dependent, and therefore only site independent forces can be reliably extracted. As the tip's oscillation amplitude was large, Durig's inversion procedure could be applied to calculate the force at the distance of closest approach from the measured frequency shift, and hence to generate the force-distance curves [22].

In contrast to the case of a flat surface, for nanotubes, either individual or bundled in a rope, the tip's interaction area changes continuously when crossing them. This can lead to contrast in the frequency images unrelated to variations in the physical properties, but entirely of topographic origin. Therefore it was essential to consider cases that allowed one to separate topographic from material contributions. Consequently, the behaviour of the tip across an individual, unfilled single walled carbon nanotube was taken here as the reference for topographic contrast.

### 3. Results

Figure 1 shows data for this reference case: (a) and (b) show images constructed from the frequency-distance and dissipation-distance curves, respectively, at –0.5 nm below the topographic profile, while (c) shows force curves taken at representative points across the nanotube and on the flat graphite surface. Here the topographic profile was featureless and symmetric, while the frequency image 1(a) is also symmetric, showing that the tip did not introduce artefacts. The frequency image 1(a) shows what we call typical topographic contrast: well defined dark regions at the edges of the nanotube (corresponding to the strongest interaction), which continuously evolve into increasingly whiter regions moving towards the top of the nanotube (corresponding to lower interaction), while the flat surface has an intermediary intensity. This behaviour is reflected also in the overall steepness of the force-distance curves: steepest at the edges, shallowest on top of the nanotube, while the curve corresponding to the flat graphite surface lies between the two. The corresponding dissipation map 1(b) appears as the mirror of 1(a), with the highest (lowest) dissipation for the steepest (shallowest) force-distance slope.

Figure 2(a)-(f) show similar images and curves as those in Figure 1, this time taken on a thin bundle of the filled material. Measurements were repeated several times showing that the tip has not changed in the process. The bundle is compact as shown by the quite symmetric line profile obtained from imaging with a feed-back set point of – 0.3 nN (corresponding to –20 Hz) (Fig. 2(a,b)). The fact that this topographic profile is rather smooth, without revealing the individual tubes of the bundle, is due to imaging with a high set-point (i.e. low interaction). There are several frequency "snap-shots", corresponding to distances above and bellow the topographic profile (Fig.2c-f). The frequency image

corresponding to –0.45 nm below the topographic profile (Fig. 2c) retrieves the same general behaviour as in Fig.1(a), with dark regions around the edges of the bundle and whiter regions towards its middle. In addition to this, in this case, faint white stripes appear on the bundle body, roughly aligned with the bundle axis, suggestive of the individual nanotubes in the bundle. This contrast is inverted in the frequency images corresponding to distances above the topographic profile (Fig. 2(e)): now the edges of the bundle are white, while the previously white stripes along the bundle body become dark.

Non-topographic contrast occurs in the region marked with "f" in the right part of the nanotube bundle. Here the overall slope of the force-distance dependence is at its lowest, resulting in a distinctively low interaction force (Fig. 3(a)). This is obtained despite region "f" being slightly on the side of the bundle, in which case the interaction should increase relative to the top of the bundle (where the interaction area, and thus the force, should be at its minimum, as shown also in Fig. 1). Consequently, this stripe shows the most pronounced contrast in all the frequency images. The dissipation map (Fig. 2(d)) mirrors the frequency map (Fig. 2(c)), same as for the single nanotube in Fig. 1, with the exception of the stripe "f". In this region there is higher dissipation than in the surroundings despite the lowest interaction (compare Fig 2(c) and (d)). This overall behaviour is highlighted also by the selected force-distance and dissipation-distance curves from Fig. 3(a) and (b). All these characteristics will be analysed in terms of material-induced contrast in section 4.

An even more complex behaviour is obtained over a nanotube bundle that is not as compact as the one described above. Based on both frequency snap-shot maps (Fig. 4(a),(c)) and force-distance curves (Fig. 4(f)), there are regions, marked with "*", for which the force-distance dependency is more complex. In these regions, the upper part of the force-distance curves have the lowest slopes of all the curves, reminiscent of the stripe "f" on the compact bundle from Fig. 2. These curves then evolve into stronger slopes when closer to the surface, as obtained on the surrounding "normal" regions. This complex behaviour suggests occurrence of multiple interactions, first through the end of the tip with a region similar to stripe "f" from Fig. 2, and then through the rest of the tip's body with adjacent parts of the bundle.

### 4. Analysis and discussion

In order to isolate and then understand material-related contrast in the measurements above, one needs first to analyse the data containing topographic information only and, by so, to provide a framework for analysing the various interaction regimes in the force-distance curves. From Fig. 1, which is the chosen reference for topographic contrast, the non-monotonic variation of the slope of the force-distance dependence across the nanotube suggests an interaction scaling with the probed area, so that relative to the flat graphite surface the interaction increases at the edges of the tube and then progressively decreases to reach a minimum on the top of the tube. A long range, additive interaction such as van der Waals would be an *a priori* candidate to explain this scaling. It is a common feature of all sets of curves, from Fig.1-4, to be best described by a tip with two parts: a mesoscopic body that is separated by a relatively large distance, about 1-2 nm, from the very end of the tip, which can be considered as a nanotip. This has been obtained from the asymptotic behaviour of the macroscopic van der Waals interaction used to fit the long-range, i.e. slowest varying part of the curves [23,13].

To quantify the contribution of this mesoscopic part of the van der Waals interaction, produced by the tip's mesoscopic body that has a relatively large offset, we proceeded in an inverse way, by simulating how it should scale when the tip crosses the individual unfilled single-walled nanotube, while keeping the short-range interaction constant. To retain only topographic variations, the physical parameters for the tube were chosen to be the same as for the graphite substrate, and as extracted from the curves in Fig. 1. The action of the feed-back has been described by calculating the tip position from the condition of constant interaction at any point above the surface, and taking as a zero-line the contact profile, obtained through the convolution of the nanotube by the tip in contact with the nanotube. Fig. 5 shows the resulting simulated force-distance curves for the case of a nanotube of 1.4 nm diameter probed by a tip terminated by a mesoscopic sphere of 11 nm radius (as de-convoluted from the topographic profile from Fig. 1) [24]. Note that the simulation took into account the fact that the tip oscillating vertically does not follow the radial symmetry of the force field. Though the maximum of the interaction scales in the expected way with the lateral position of the tip, i.e. being the highest at the very edges of the tube and the smallest on the top of the tube, the order in which the force curves disperse around the feed-back set-point (i.e. the point where they intersect) does not correspond to the experimental case (compare Fig. 5(a) and Fig. 3(a)). Increasing the feed-back set-point to higher force even eliminates this dispersion (Fig. 5(b)), which would lead to no contrast at all in the frequency maps. Consequently, the simulation suggests that in our experiments the mesoscopic van der Waals force alone, because of the large distance offset between the meso- and the nano- parts of the tip, has only a weak contribution in determining the trend of the force-distance curves when the tip moves across the nanotube(s).

The short-range part of the interaction, that remains after extraction of the long –range van der Waals contribution varies significantly, both in magnitude and overall extent (of the order of 1nm), when the tip moves from the flat surface across the nanotube (or bundle of nanotubes), and is therefore responsible for the contrast seen in the frequency "snap-shot" maps. The characteristic length of this interaction $\lambda$ [25] is large, of about 0.3 nm on graphite surface and on various locations on the nanotubes, and increases to about 0.5 nm in the region "f" of the nanotube bundle from Fig. 2. These large $\lambda$ values indicate that the dominant interaction involved is certainly not chemical bonding (as this should be truly short-ranged, about 2-3 Å overall extent [14]). Chemical bonding is unlikely to occur as both graphite and nanotubes (even when filled) are non-reactive surfaces, while the tip is originally $SiO_2$, i.e. without dangling bonds. Instead, a likely contributor is the van der Waals interaction between nano-scale objects. Unlike the van der Waals interaction between mesoscopic bodies which is long-ranged, if the interacting objects are nano-sized the distance dependence of the interaction can be much stronger, and thus of a shorter range [26]. Accordingly, by approximating the nano-tip by a small sphere of radius $a$, its interaction force with the flat surface varies as $1/D^4$ for $a \leq D$, and tends to $1/D^2$ very close to the surface, for $a >> D$ (when the small sphere appears as macroscopic relative to $D$). Similarly, a sphere-sphere interaction force changes from $1/D^7$ to $1/D^2$ with decreasing separation $D$. The sphere-cylinder interaction lies between these two cases. Thus, a "nano-scale" van der Waals interaction can account for a force-distance dependency according to a power law whose exponent, and thus extent, varies with the tip's lateral position across the nanotube, but also when approaching the surface, as observed in our measurements. Moreover, the change in the strength of the interaction (for example, from sphere-plane to sphere-cylinder) appears naturally in this framework.

It is also possible that the extent of the short-range interaction is increased by a residual electrostatic interaction, originating from uncompensated contact potential differences over the nanotubes compared to the graphite substrate. Though contact potential measurements were not performed simultaneously with the frequency maps, separate measurements showed systematic work function variations of about 0.2 eV on samples rich in Ag@SWNT nano-composites [27]. Electron transfer from Ag to the carbon nanotube was also indicated by Raman spectroscopy on bulk SWNTs modified by a combination of filling and intercalation [9].

We focus now on the aspects related to material contrast as obtained over region "f" in Fig. 2, i.e. a lower force, with longer characteristic length, and higher dissipation, and which we tentatively attribute to presence of the Ag filling. Several possible sources of contrast can be involved.

Firstly, we consider the changes in polarizability due to Ag filling of the nanotube. According to McLachlan-Lifshitz susceptibility theory, the van der Waals dispersion interaction $w$ between two small particles in a medium depends on the product of polarizabilities of the two particles in that medium, and, therefore, on the dielectric properties of the media involved [28]. In our case, the interacting particles are the nanometer-sized end of the tip, and the filled/unfilled nanotubes. $w$ has two contributions, a static one $w_{\nu=0}$, and a dynamic one $w_{\nu>0}$, which for the case of two spherical particles take the form:

$$w_{\nu=0} = -\frac{A_{H\nu=0}a_1^3 a_2^3}{D^n} = -\frac{3kTa_1^3 a_2^3}{D^n}\left(\frac{\varepsilon_1(0)-\varepsilon_3(0)}{\varepsilon_1(0)+2\varepsilon_3(0)}\right)\left(\frac{\varepsilon_2(0)-\varepsilon_3(0)}{\varepsilon_2(0)+2\varepsilon_3(0)}\right) \qquad (1a)$$

$$w_{\nu>0} = -\frac{A_{H\nu>0}a_1^3 a_2^3}{D^n} = -\frac{3ha_1^3 a_2^3}{D^n}\int_0^\infty\left(\frac{\varepsilon_1(i\nu)-\varepsilon_3(i\nu)}{\varepsilon_1(i\nu)+2\varepsilon_3(i\nu)}\right)\left(\frac{\varepsilon_2(i\nu)-\varepsilon_3(i\nu)}{\varepsilon_2(i\nu)+2\varepsilon_3(i\nu)}\right)d\nu \qquad (1b)$$

where $\varepsilon(i\nu)$ are generated from functionals (usually, sums of quantum mechanical oscillators) describing the corresponding complex dielectric functions [29], by replacing $\nu$ by $i\nu$ [30]. The exponent $n$ describes a generic distance dependence. Here media "1", "2" and "3" refer to the tip, the nanotube (filled or not), and the embedding vacuum (i.e. $\varepsilon_3(i\nu) = 1$), respectively. Referring to the medium "2", there are two cases to be compared, (i) that of the unfilled nanotube when $\varepsilon_2 = \varepsilon_{SWNT}$, and (ii) the one of the filled nanotube, when $\varepsilon_2 = \varepsilon_{SWNT/Ag}$ is the effective dielectric function of the composite system Ag filling-SWNT shell [31]. Due to the complexity of the calculations, we evaluated sphere-sphere [28] instead of sphere-cylinder interactions, while the anisotropic SWNT was described by the transverse dielectric function $\varepsilon_{SWNT}^{\perp}$ of a (23,0) semiconducting nanotube [32]. This last simplification is justified by the fact that the interaction mediated by the tip is transverse to the nanotube, and so is the depolarization field created by the bound charge that builds up on the nanocomposite's surface. Unlike $\varepsilon_{SWNT}^{\parallel}$, $\varepsilon_{SWNT}^{\perp}$ is not sensitive to the semiconducting or metallic nature of the nanotube [33], but only moderately dependent on the tube's diameter [33], therefore one does expect the evaluations performed here based on a specific choice of nanotube shell to have a more general significance. Figure 6 reveals some notable facts: (i) though both Ag [34] and SWNT [32] can reach large dielectric function values, because of their opposite signs their combination gives an effective dielectric function $\varepsilon_{SWNT/Ag}$ for the Ag filling-SWNT shell system that varies only slightly (Fig. 6(b)). (ii) Functions $\varepsilon(i\nu)$ are positive and have monotonic variations for all the systems involved (Fig. 6(c)), despite sharp peaks in the originating dielectric functions (such as for $\varepsilon_{SWNT}$); this is due to the fact that the substitution $\nu \rightarrow i\nu$ in the sum of oscillators form of $\varepsilon$ removes any existing poles for positive and real values of $\nu$, except potentially at $\nu = 0$. Note also that $\varepsilon(i\nu)$ for the independent Ag and SWNT systems exceed by far that of the combined SWNT/Ag system, behaviour that seems to follow the overall variations that are reached in each case by the originating dielectric functions. As the factors under the integrals in relation 1(b) are all positive ($\varepsilon(i\nu)$ are all higher than unity), the resulting Hamaker constants (shown in Table I) reflect the same trend as the $\varepsilon(i\nu)$ functions, being higher when the tip interacts with the unfilled SWNT than for the Ag filling-SWNT shell system. This calculation thus supports our experimental observations. Table I shows also that replacing the $SiO_2$ tip with a metallic one (for example, of Ag) does not change this trend, but only increases the strength of the van der Waals interaction in all cases.

Dissipation-distance curves (Fig. 3(b)) reveal through better defined slope changes that there are several interaction regimes when decreasing the tip's average separation to the surface. This occurs in a clearer way than for the force-distance curves (Fig. 3(a)), where the slope changes are more gradual. The dissipation curves have a similar shape, whether over the flat graphite or on different points across the nanotube rope, with three apparent regimes: (1) at large tip-sample separations, where mesoscopic van der Waals and electrostatic forces dominate, (2) an intermediary regime controlled by the shorter range dispersion forces with $1/D^n$ dependency, as discussed above, and (3) a third regime at the lowest separations corresponding to the force's $1/D^n$ to $1/D^2$ transition, and where repulsive non-contact forces potentially also start to contribute. We believe that in the curves shown in Figure 3 a contact repulsive regime is not reached, as this would be characterized by a much steeper increase in the dissipation (data not shown here). In addition, for relatively hard materials, such as nanotubes and graphite, one expects the onset of the repulsive regime to be signalled by a clear minimum in the force-distance curves [12].

Energy dissipation can occur through the visco-elastic response of the surface to the interaction with the tip [35] (for a self-driven cantilever there is no additional apparent damping due to non-dissipative processes [36]). This response is described by the mechanical susceptibility $S(\omega) = \omega\gamma/\left(k_s^2 + \gamma^2\omega^2\right)$, where $k_s$ is the surface's spring constant, while $\gamma_s$ is a viscosity related damping factor. For hard surfaces such as nanotubes, one expects a fast, elastic response of the surface (i.e. the sample deformation follows the tip with little phase delay) rather than a frictional one [37]. This regime occurs when the tip's residence time close to the surface $\tau_{res} \approx \left(1/\pi f_0\right)\left(2D/A\right)^{1/2}$ is longer than the surface's visco-elastic characteristic time $\gamma_s/k_s$ [37]. For our experimental conditions, $\tau_{res} \approx 10^{-7}$ s, while $\gamma_s/k_s \approx 10^{-10}$ s using $k_s \approx 2$ nN/Å, estimated from asymmetric compression of the SWNT wall [38], and $\gamma_s \approx 10^{-9}$ Ns/m, which is an overestimation for this parameter, corresponding to a block copolymer (a softer material) [39]. In this case, the dissipation caused by an attractive force falling as $1/D^n$ varies as:

$$\frac{D}{D_\infty} - 1 = \frac{A_{exc}}{A_{eex\infty}} - 1 \propto \left(\frac{A_H a}{D^n}\right)^2 \frac{1}{k_s}\frac{1}{A^2} \qquad\qquad \left(\tau_{res} >> \frac{\gamma}{k_s}\right) \qquad (2)$$

Dissipation is now phonon assisted, and controlled by the surface's elasticity. Under this assumption, the "odd" behaviour over the region "f", where dissipation increases despite a decrease in the interaction strength, can be explained according to relation (2) by an increase in elasticity (i.e. lower $k_s$) compared to nearby regions. Fitting the lower parts of the dissipation-distance curves shows that dissipation follows roughly the square of our van der Waals dispersion forces, though a very precise determination of the power exponent and, thus, firm assertion of this dissipation regime is not possible at this stage.

That the SWNT's elasticity, controlling phonon-assisted dissipation, changes as a result of filling is supported by several pieces of evidence. Significant relaxation and even deformation of the SWNT shell have been observed or predicted for various types of fillings. Raman spectroscopy has shown a down-shift of the G peak, consistent with softening of the C-C bonds, in Ag filled nanotubes [9], while HRTEM observations on the RbI@SWNT system revealed an extreme case, where the SWNT shell has changed shape, from cylindrical to slightly squared [40]. Theoretical calculations on the KI@SWNT also showed that relaxation of about 0.05 Å of the C atoms positions is expected in this case [41]. All these examples support the idea that, for a number of filling materials, the nanotube shell can become more deformable in the limit of very small deformations (i.e. of the order of the relaxation distance), as would be encountered in non-contact dissipation.

In conclusion, we have presented a force spectroscopy study on filled carbon nanotube systems. The Ag@SWNT system appears as appropriately chosen to reveal contrast based on the van der Waals dispersion interaction and energy dissipation. These were linked to material properties, such as effective dielectric function of the nano-composite and elasticity of the relaxed carbon nanotube shell. From this starting point, more specific and quantitative information can be obtained in a more controlled experiment, where measurements are taken on a partially filled individual SWNT, instead of a bundle. Modelling of the dielectric properties based on the specific structure (as obtained by HRTEM) and the one-dimensional nature of the Ag filling is necessary for a correlated understanding of the effects.

### Acknowledgments

This work was supported by the Interdisciplinary Research Collaboration in Nanotechnology, at Cambridge University. A.I. gratefully acknowledges Alex Shluger for useful discussions, and Andrey Gal for sharing some of his unpublished theoretical results.

**References**

[1] B.W. Smith, M. Monthioux, and D.E. Luzzi, *Nature* **396**, 323 (1998); B.W. Smith, D.E. Luzzi, and Y. Achiba, *Chem. Phys. Lett.* **331**, 137 (2001); K. Suenaga, M. Tence, C. Mory. C. Colliex, H. Kato, T. Okazaki, H. Shinohara, K. Hirahara, S. Bandow, and S. Iijima, *Science* **290**, 2280 (2000).

[2] R.. Meyer, J. Sloan, R.E. Dunin-Borkowski, A.I. Kirkland, M.C. Novotny, S.R. Bailey, J.L. Hutchinson, M.L.H. Green, *Science* **289**, 1324 (2000).

[3] J. Sloan, A.I. Kirkland, J.L. Hutchinson, and M.L.H. Green, *Chem. Comm.* **13**, 1319 (2002).

[4] D.A. Morgan, J. Sloan, and M.L.H. Green, *Chem. Comm.* **20**, 2442 (2002).

[5] T. Takenobu, T.Takano, M. Shiraishi, Y. Murakami, M. Ata, H. Kataura, Y. Achiba, and Y. Iwasa, *Nature Mat.* **2**, 683 (2003).

[6] Y. Xia, P. Yang, Y. Sun, Y. Wu, B. Mayers, B. Gates, Y. Yin, F. Kim, and H. Yan, *Adv. Mat.* **15**, 353 (2003), and references therein.

[7] M. Nishizawa, V.P. Menon, C.R. Martin, *Science* **268**, 700 (1995); B.H. Hong, S.C. Bae, C. Lee, S. Jeong, K.S. Kim, *Science* **294**, 348 (2001); S.J. Limmer, S. Seraji, Y. Wu, T.P. Chou, C. Nguyen, G. Cao, *Adv. Func. Mater.* **12**, 59 (2002); M. Barbic, J.J. Mock, D.R. Smith, S. Schultz, *J. Appl. Phys.* **91**, 9341 (2002).

[8] R. Carter, unpublished.

[9] P. Corio, A.P. Santos, P.S. Santos, M.L.A. Temperini, V.W. Brar, M.A. Pimenta, and M.S. Dresselhaus, *Chem. Phys. Lett.* **383**, 475 (2004).

[10] S. Suzuki, F. Maeda, Y. Watanabe, T. Ogino, *Phys. Rev. B* **67**, 115418 (2003).

[11] F.X. Zha, R. Czerw, D.L. Carroll, P. Kohler-Redlich, B.Q. Wei, A. Loiseau, S. Roth, *Phys. Rev. B* **61**, 4884 (2000); R. Czerw, J.W. Liu, D.L. Carroll, *New Jour. Phys.* **6**, art. No. 031 (2004).

[12] H. Holscher, A. Schwarz, W. Allers, U.D. Schwarz, and R. Wiesendanger, *Phys. Rev. B* **61**, 12678 (2000);

[13] M. Guggisberg, M. Bammerlin, Ch. Loppacher, O. Pfeiffer, A. Abdurixit, V. Barwich, R. Benewitz, A. Baratoff, E. Mayer, and H.-J. Guntherodt, *Phys. Rev. B* **61**, 11151 (2000).

[14] M.A. Lantz, H.J. Hug, R. Hoffmann, P.J.A. van Schendel, P. Kappenberger, S. Martin, A. Baratoff, A. Abdurixit, H.-J. Guntherodt, *Science* **291**, 2580 (2001); R. Hoffmann, L.N. Kantorovich, A. Baratoff, H.J. Hug, and H.J. Guntherodt, *Phys. Rev. Lett.* **92**, 146103 (2004).

[15] H.W.C. Postma, A. Sellmeijer, C. Dekker, *Adv. Mat.* **12**, 1299 (2000).

[16] T.W. Tombler et al., *Nature* **405**, 769 (2000); E.D. Minot, Y. Yaish, V. Sazonova, J.-Y. Park, M. Brink, P.L. McEuen, *Phys. Rev. Lett.* **90**, 156401 (2003).

[17] S.J. Tans, C. Dekker, *Nature* **404**, 834 (2000).

[18] A. Bachtold, M.S. Fuhrer, S. Plyasunov, M. Forero, E.H. Anderson, A. Zettl, P.L. McEuen, *Phys. Rev. Lett.* **84**, 6082 (2000).

[19] B. Gotsmann, C. Seidel, B. Anczykowski, and H. Fuchs, *Phys. Rev. B* **60**, 11051 (1999); C. Loppacher, R. Bennewitz, O. Pfeiffer, M. Guggisberg, M. Bammerlin, S. Schar, V. Barwich, A. Baratoff, and E. Meyer, *Phys. Rev.B* 62, 13674 (2000).

[20] J. Sloan, D.M. Wright, H.-G. Woo, S. Bailey, G. Brown, A.P.E. York, K.S. Coleman, J.L. Hutchinson, and M.L.H. Green, *Chem Comm.* 699 (1999).

[21] For amplitudes that are large compared to the range of the interaction force, long range forces have an important contribution to the frequency shift, as described in F.J. Giessibl, H Bielefeldt, *Phys. Rev. B* **61**, 9968 (2000). In this regime, the frequency shift can be rendered independent of the oscillation conditions by normalization to $f_0 / \left( k_c A^{3/2} \right)$.

[22] U. Durig, *Appl. Phys. Lett.* **76**, 1203 (2000).

[23] R. Hoffmann, M.A. Lantz, H.J. Hug, P.J.A. van Schendel, P. Kappenberger, S. Martin, A. Baratoff, and H.-J. Guntherodt, *Phys. Rev. B* **67**, 085402 (2003).

[24] Despite the presence of the nano-tip, the mesoscopic body of the tip still contributes to enlarging the topographic profile of the nanotube.

[25] For the purpose of comparing force-distance curves only, the short-range part of the interaction was described empirically in the form of a Morse potential, with both attractive and repulsive parts:

$$F = 2U_0 / \lambda \; \left\{ \exp\left[ -2(D - D_0)/\lambda \right] - \exp\left[ -(D - D_0)/\lambda \right] \right\}.$$ No specific physical basis is attached to this description.

[26] J.N. Israelachvili, *"Intermolecular and Surface Forces",* chapter 10, p. 152-175, Academic Press (1992).

[27] A. Ilie, unpublished.

[28] A.D. McLachlan, *Discuss. Faraday Soc.* **40**, 239 (1965); E.M. Lifshitz, *Soviet Phys. JETP (Engl. Transl.)* **2**, 73 (1956).

[29] V.A. Parsegian, G.H. Weiss, *J. Coll. Inter. Sci.* **81**, 285 (1981).

[30] D.B. Hough, L.R. White, *Adv. Coll. Inter. Sci.* **14**, 3 (1980).

[31] The effective dielectric function has been calculated based on a simplified spherical core-shell model, as in C.F. Bohren and D.R. Huffman, *"Absorption and Scattering of Light by Small Particles"* (John Wiley and Sons, Inc., 1998), chapter 5, pp. 130-154. This model required the fraction $\nu$ of the total volume of the core-shell particle occupied by the Ag core. A wall thickness of the single-walled nanotube of 0.67 Å, and a spacing of 3.4 Å between the shell and the Ag filling were used for the estimation of $\nu$.

[32] M.F. Lin, K. W.-K. Shung, *Phys. Rev. B* **50**, 17744 (1994).

[33] L.X. Benedict, S.G. Louie, M.L. Cohen, Phys. Rev. B **52**, 8541 (1995).

[34] P.B. Johnson, R.W. Christy, *Phys. Rev. B* **6**, 4370 (1972).

[35] J.P. Aime, D. Michel, R. Boisgard, and L. Nony, *Phys. Rev. B* **59**, 2407 (1999).

[36] H. Holscher, B. Gotsmann, W. Allers, U.D. Schwarz, H. Fuchs, and R. Wiesendanger, *Phys. Rev. Lett.* **88**, 019601 (2002).

[37] G. Couturier, J. P. Aime, J. Salardenne, R. Boisgard, A. Gourdon, and S. Gauthier, *Appl. Phys. A* **72**, S47 (2001).

[38] V. Lordi, N. Yao, *J. Chem. Physics* **109**, 2509 (1998).

[39] F. Dubourg, S. Kopp-Marsaudon, Ph. Leclere, R. Lazzaroni, and J.P. Aime, *Eur. Phys. J. E* **6**, 387 (2001).

[40] M.L.H. Green, private communication.

[41] A. Gal, private communication.

**Figure captions**

**Figure 1**

Typical spectroscopic mapping and curves over an individual, unfilled single-walled nanotube. (a) and (b) frequency and dissipation maps ($0.6 \times 1.6 \, nm^2$ resolution), respectively, corresponding to a tip's position -0.5 nm below the topographic profile. (c) Representative force-distance curves for various lateral positions of the tip, on the flat surface and across the nanotube. The curves are symmetric relative to the centre of the tube's topographic profile. The strength of the interaction decreases from the sides towards the top of the nanotube, where it reaches its minimum.

**Figure 2**

Spectroscopic maps ($1 \times 1 \, nm^2$ resolution) and topography of a bundle from the Ag filled material. (a) and (b) topography and topographic profile, respectively, taken with a set point of −0.3 nN (corresponding to −20 Hz in the original frequency curves). (c) and (d) frequency and dissipation maps corresponding to the tip being positioned −0.45 nm below the topographic profile from (a). (e) and (f) frequency and dissipation maps corresponding to the tip being positioned above the topographic profile from (a). Region "f", with lowest force and higher dissipation, is attributed to an Ag filled part of a nanotube.

**Figure 3**

Selected force-distance (a) and dissipation-distance (b) curves corresponding to the bundle from Fig. 2, taken at various lateral positions of the tip: on the flat graphite surface (curve 1), and across the nanotube bundle (2 at the very edges of the bundle, 3 at the very top of the bundle, and "f" over the region "f"). Curves with stronger interaction force correspond to curves with stronger dissipation, except for region "f", where dissipation is not the lowest despite the force being the lowest. The force values are compatible with nano-scale van der Waals interaction between one nanometer radius sphere and tube/plane.

**Figure 4**

Frequency maps (a) and (c), and dissipation maps (b) and (d) below and above the topographic profile, respectively. $1 \times 1 \, nm^2$ resolution. The marked regions "*" from (c) are attributed to Ag filled nano-composites. (e) is the superposition of (a) and (c). (f) Force –distance curves corresponding to a Ag filled region ("*"), the edge of the bundle (1), and on a normal region on the bundle (2).

**Figure 5**

(a) Simulated van der Waals interaction between a mesoscopic tip, of 11 nm radius, and the single-walled nanotube from Figure 1. Curves correspond to different points across the topographic profile shown in the inset. The inset shows the contact profile (continuous line) and the topographic profile (dashed line) obtained for a given set-point. (b) corresponds to a lower set-point (higher force) for the feed-back.

**Figure 6**

(a) Complex dielectric functions for Ag, and corresponding to the transverse direction for a (23,0) SWNT [32]. (b) Effective dielectric function for the Ag filling-SWNT shell system. A spherical system was considered instead of the real cylindrical geometry. (c) $\varepsilon(i\nu)$ functions for Ag, unfilled SWNT, and the combined Ag/SWNT system.

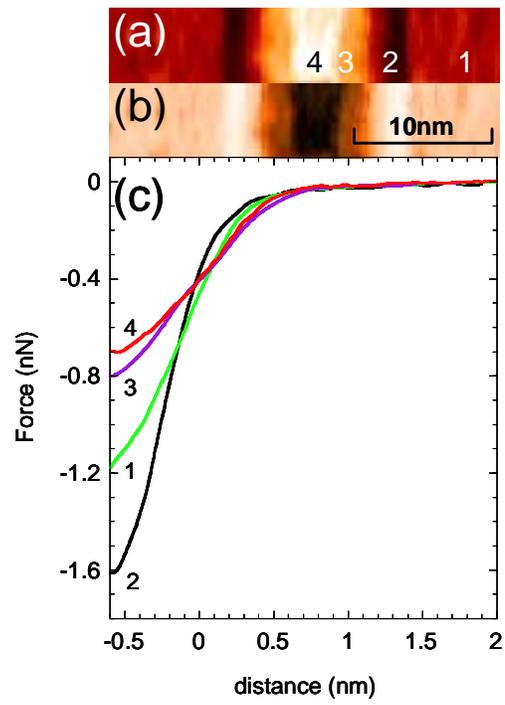

Figure 1

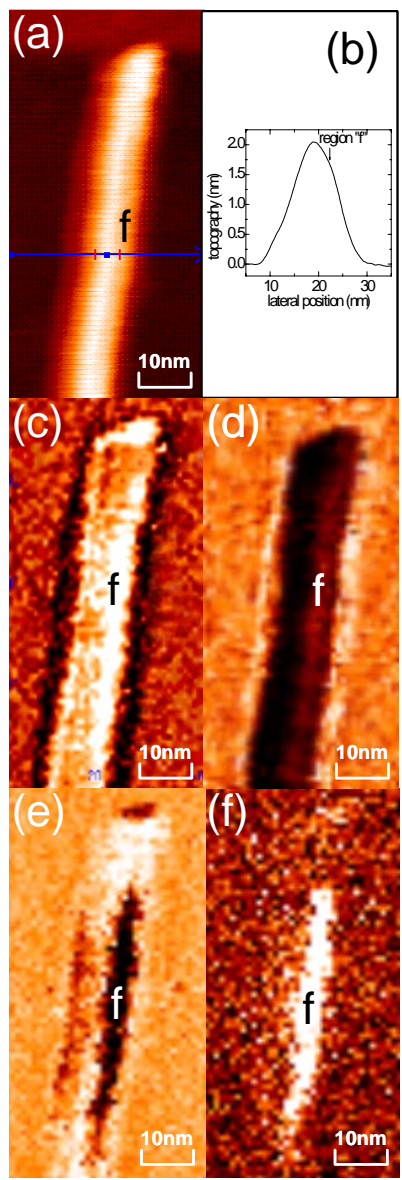

Figure 2

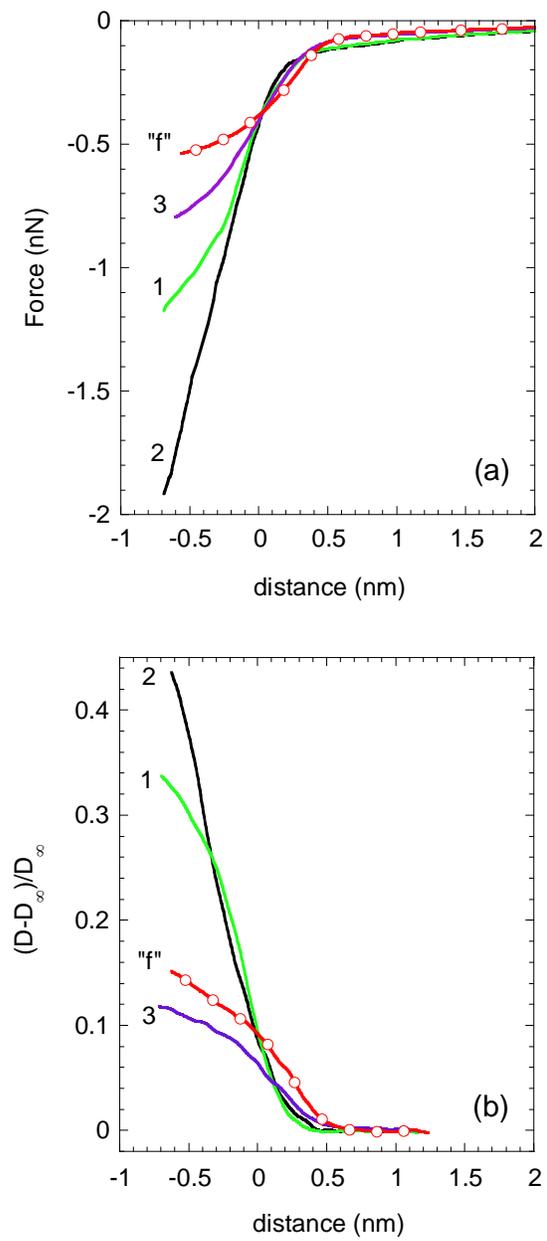

Figure 3

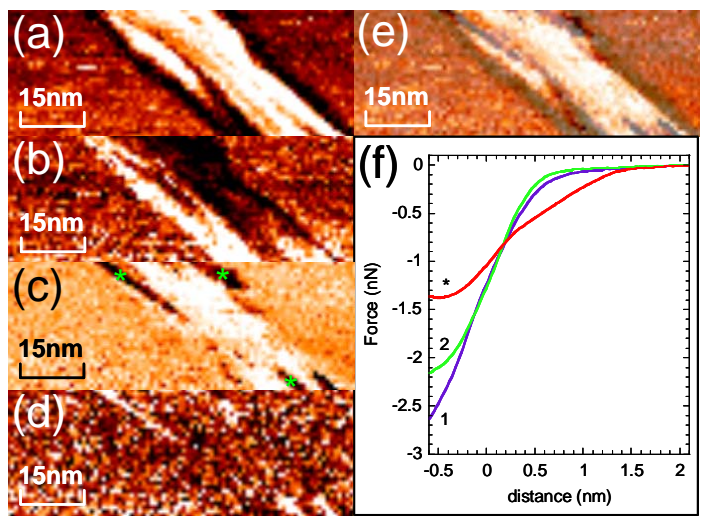

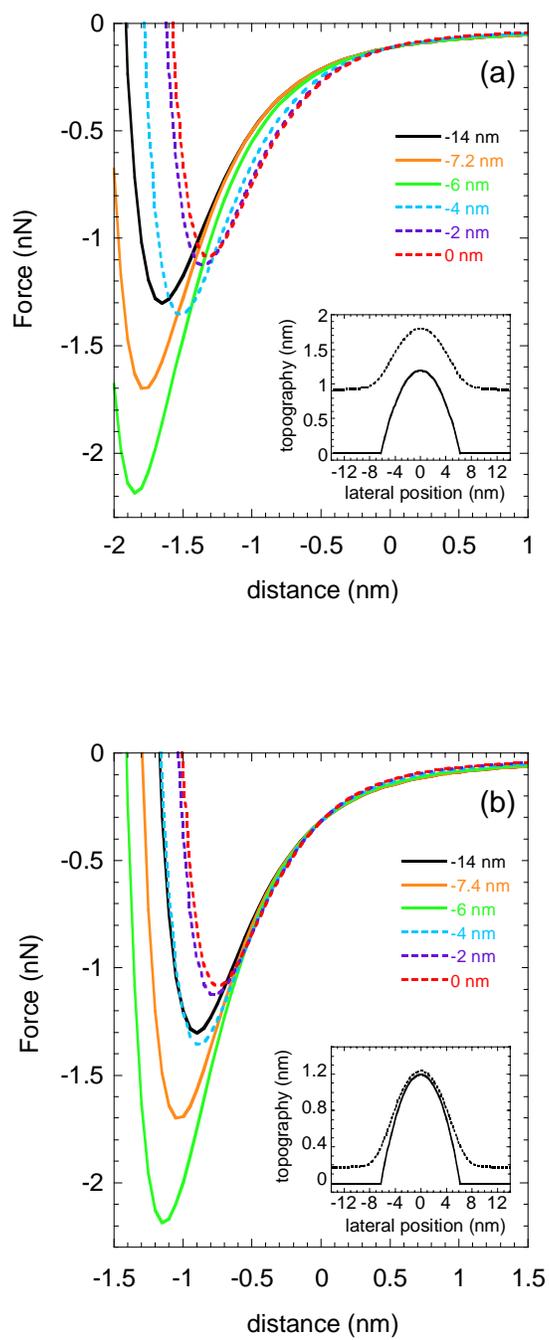

Figure 5

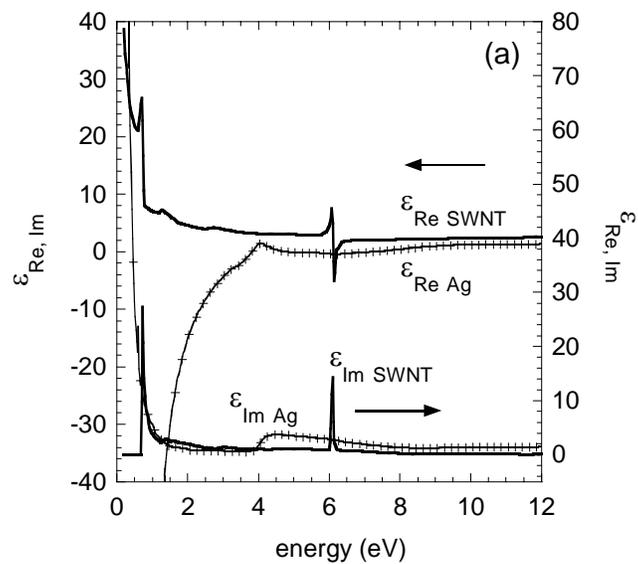

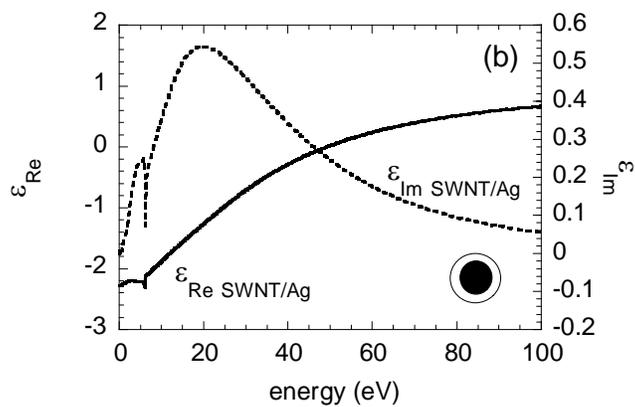

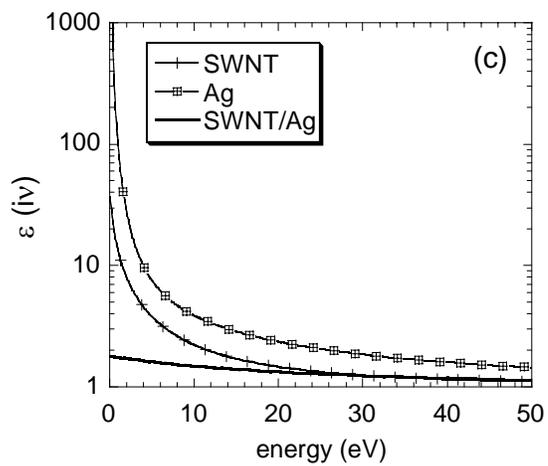

Figure 6

| Hamaker constant ($\times 10^{-20}$ J) | SiO$_2$ tip | | | Ag tip | | |
|---|---|---|---|---|---|---|
| | Core-shell (Ag-SWNT) particle | Unfilled graphitic shell | Ag particle | Core-shell (Ag-SWNT) particle | Unfilled graphitic shell | Ag particle |
| $A_{H\,v=0}$ | 0.03 | 0.15 | 0.16 | 0.06 | 0.29 | 0.31 |
| $A_{H\,v>0}$ | 3.27 | 7.75 | 12.84 | 8.64 | 20.71 | 32.69 |
| $A_H$ | 3.3 | 7.9 | 13 | 8.7 | 21 | 33 |

Table I

Hamaker constants for various interacting systems, comparing static and dynamic components.